\documentstyle[12pt,a4,epsf]{article}

\makeatletter
\@addtoreset{equation}{section}
\makeatother

\newcommand{\VEV}[1]{\left\langle #1 \right\rangle}

\newcommand{\bequ}{\begin{equation}}
\newcommand{\eequ}{\end{equation}}
\newcommand{\beqn}{\begin{eqnarray}}
\newcommand{\eeqn}{\end{eqnarray}}
\newcommand{\bctr}{\begin{center}}
\newcommand{\ectr}{\end{center}}

\begin{document}
\begin{titlepage}

\begin{flushright}
hep-ph/0310232\\
KUNS-1877\\
\today
\end{flushright}

\vspace{4ex}

\begin{center}
{\large \bf
GUT with Anomalous $U(1)_A$ Suggests Heterotic M-Theory?
}

\vspace{6ex}

\renewcommand{\thefootnote}{\alph{footnote}}
Nobuhiro MAEKAWA\footnote
{E-mail: maekawa@gauge.scphys.kyoto-u.ac.jp
}

\vspace{4ex}
{\it Department of Physics, Kyoto University, Kyoto 606-8502, Japan}\\
\end{center}

\renewcommand{\thefootnote}{\arabic{footnote}}
\setcounter{footnote}{0}
\vspace{6ex}

\begin{abstract}
We show that a new GUT scenario we proposed
indicates that heterotic M-theory is one of the most interesting
possibility to describe our world because of the two features in
the scenario.
The first feature is that $E_6$ unified group plays an essential
role not only in understanding larger mixing angles in lepton sector
 than in quark sector but also in solving the SUSY flavor problem 
 by introducing non-abelian horizontal symmetry
 $SU(2)_H$ or $SU(3)_H$.
The second feature is that the cutoff scale must be taken as
the usual GUT scale $\Lambda_G\sim 2\times 10^{16}$ GeV to realize
natural gauge coupling unification. Because this cutoff scale is 
smaller than the Planck scale or string scale, it may suggest the 
existence of extra dimension in which only gravity modes can propagate.
This talk is based on a dozen 
papars\cite{maekawa,BM,E6Higgs,sliding,NGCU,maekawa2,horiz,flipped,shafi}
some of which are collaborated
with T. Yamashita, M. Bando and Q. Shafi.

\end{abstract}

\end{titlepage}

\section{Introduction and Overview of our scenario}
One of the biggest difference between our GUT scenario and various 
previous GUT models is to introduce generic interactions (including higher
dimensional interactions) with $O(1)$ coefficients
in our scenario. Therefore, once the symmetry is fixed, the theory can
be defined except the $O(1)$ coefficients. It is also quite amazing that
almost all the problems can be solved with generic interactions;
realistic quark and lepton masses and mixings (predicting LMA solution
for the solar neutrino problem) can be obtained\cite{maekawa,BM};
doublet-triplet splitting can be realized without too rapid proton
decay via dimension five operators\cite{maekawa,E6Higgs,sliding};
this scenario gives a new natural explanation for the success of gauge
coupling unification in the minimal SUSY standard model(MSSM) although
$E_6$ gauge group, whose rank is larger than that of $SU(5)$, is adopted 
as the unified gauge group\cite{NGCU}; if SUSY breaking sector is introduced,
$\mu$ problem is also solved\cite{maekawa2}; 
if non-abelian horizontal gauge symmetry
$SU(2)_H$ or $SU(3)_H$ is introduced, SUSY flavor problem is also 
solved.\cite{horiz}

It is obvious that introducing generic interactions simplifies the process
to obtain a realistic model from the superstring theory, because
a symmetry defines a model.
Let me show an example.
Just the symmetry is presented in Table 1.
\begin{table}[t]
\caption{Quantum numbers under $E_6\times SU(2)_H\times U(1)_A\times Z_2$
are presented only for all non-singlet fields under $E_6\times SU(2)_H$ except
an doublet field $X(1,2)_x$ which is required for Witten anomaly cancellation.
\label{su2}}
\vspace{0.2cm}
\begin{center}
\footnotesize
\begin{tabular}{|c|l|}
\hline
MATTER & $\Psi({\bf 27},{\bf 2})_4^+$, 
         $\Psi({\bf 27},{\bf 1})_{3/2}^+$ \\
\hline
HIGGS($SU(2)_H$) & $F({\bf 1},{\bf 2})_{-3/2}^+$, 
                   $\bar F({\bf 1},{\bf \bar 2})_{-5/2}^+$ \\
\hline
HIGGS & $A({\bf 78},{\bf 1})_{-1}^+$, $A'({\bf 78},{\bf 1})_5^+$ \\
$E_6\rightarrow $ & $\Phi({\bf 27},{\bf 1})_{-3}^+$, 
                    $C'({\bf 27},{\bf 1})_6^-$ \\
$SU(3)_C\times U(1)_{EM}$ & $\bar \Phi'({\bf \overline{27}}, {\bf 1})_5^+$,
                            $\bar C({\bf \overline{27}},{\bf 1})_0^-$ \\
\hline
\end{tabular}
\end{center}
\end{table}
Note that just two fields $\Psi$ and $\Psi_3$ include all the three 
generation matter fields. Among the eight Higgs fields, only non-primed
fields have non-vanishing vacuum expectation values (VEVs). And 
doublet Higgs in MSSM are included in $\Phi$ in this model.
If the matter and Higgs ($SU(2)_H$) sectors change as in Table \ref{su3},
all the three generation matter fields can be unified into a single
multiplet $\Psi({\bf 27},{\bf 3})$ under $E_6\times SU(3)_H$.
\begin{table}
\caption{Quantum numbers under $E_6\times SU(3)_H\times U(1)_A\times Z_2$
are presented for matter and Higgs for $SU(3)_H$. The Higgs sector for
breaking $E_6$ is the same as in Table 1.
Cancellation of $SU(3)_H$ gauge anomaly requires  some other fields, 
for example a ${\bf 10}$ of $SU(3)_H$.
\label{su3}}
\vspace{0.2cm}
\begin{center}
\footnotesize
\begin{tabular}{|c|l|}
\hline
MATTER & $\Psi({\bf 27},{\bf 3})_{7/2}^+$ \\
\hline
HIGGS($SU(3)_H$) & $F_i({\bf 1},{\bf 3})_{(-3,2)}^+$, 
                   $\bar F_i({\bf 1},{\bf \bar 3})_{(-2,-2)}^+$ $(i=2,3)$ \\
\hline
\end{tabular}
\end{center}
\end{table}


It is non-trivial that in these models, with generic interactions, 
doublet-triplet splitting is realized.\cite{E6Higgs,sliding}
Here we would like to concentrate on several points which suggest
that the heterotic M-theory may be promizing. In the scenario, $E_6$ (or
$E_8$) GUT gauge group plays an important role in understanding larger
neutrino mixing angles\cite{BM} and 
in solving SUSY flavor problem by introducing
non-abelian horizontal gauge symmetry $SU(2)_H$ or $SU(3)_H$.\cite{horiz}
It is interesting that a structure in $E_6$ models simultaneously realizes 
two features,
large neutrino mixing angles and suppressed flavor changing neutral currents
(FCNC). Because $E_6\times SU(3)_H$ is a maximal subgroup of $E_8$, the
heterotic string must be promizing. 
Moreover, the scenario of GUT with anomalous $U(1)_A$ gauge symmetry generally
explain why the three gauge couplings meet at a scale in MSSM.\cite{NGCU} 
This novel 
mechanism requires the cutoff scale around the usual GUT scale 
$\Lambda_G\sim 2\times 10^{16}$ GeV, which are smaller than the Planck scale
or the string scale. This may imply the extra dimension in which only gravity
modes can propagate as discussed by Horava-Witten.\cite{HW}

\section{Anomalous $U(1)_A$ gauge symmetry}
Anomalous $U(1)_A$ gauge symmetry is a $U(1)$ gauge symmetry with gauge anomaly
which is cancelled by Green-Schwarz mechanism.\cite{GS} But in our scenario, 
we use
anomalous $U(1)_A$ gauge symmetry as just $U(1)$ gauge symmetry with 
Fayet-Illiopoulos $D$-term whose parameter $\xi$ is smaler than the cutoff
scale $\Lambda$. If a field $\Theta$ with the negative charge $\theta=-1$ 
is introduced, 
$D$-flatness condition for $U(1)_A$ $D_A=\xi^2-|\Theta|^2=0$ determines the
VEV as $\VEV{\Theta}=\xi\equiv \lambda\Lambda$.(In this paper,
we take $\lambda\sim \sin\theta_C\sim 0.22$.)
 Therefore, the anomalous 
$U(1)_A$ gauge symmetry is broken just below the cutoff scale. 

Ibanez-Ross\cite{IR} pointed out that this field $\Theta$ can play the same role
as the Froggatt-Nielsen field\cite{FN} in SUSY models. 
Assigning the anomalous $U(1)_A$
charges for quark, lepton and Higgs superfields, gauge invariant interactions
become Yukawa interactions after developing the VEV 
$\VEV{\Theta}\sim \lambda\Lambda$ as
\begin{equation}
W=\left(\frac{\Theta}{\Lambda}\right)^{q_i+u_j+h_u}Q_iU_jH_u\rightarrow 
\lambda^{q_i+u_j+h_u}Q_iU_jH_u, (i,j=1,2,3),
\end{equation}
where lowercase letters denotes the anomalous $U(1)_A$ charge and we use 
a unit in which $\Lambda=1$.
It is obvious that larger charge leads to smaller coupling and therefore
smaller mass. And if we adopt $h_u=h_d=0$, $(q_i)=(3,2,0)$, $(u_i)=(4,2,0)$,
$(d_i)=(3,2,2)$, realistic quark masses and mixings can be obtained.
Possible criticism for this obserbation may be that any mass spectrum (6 masses
and 3 mixing) can be obtained by assigning their 9 charges.
However, this scenario generally gives Cabbibo-Kobayashi-Maskawa matrix
as $(V_{CKM})_{ij}\sim \lambda^{|q_i-q_j|}$, which leads to a non-trivial 
relation
$V_{13}\sim V_{12}V_{23}$ which is consistent with the experimental values.

How about in lepton sector? Because neutrino masses and mixings are obtained by
\begin{equation}
(m_\nu)_{ij}\sim \lambda^{l_i+l_j+2h_u}\frac{\VEV{H_u}^2}{\Lambda},
\quad (V_{MNS})_{ij}\sim \lambda^{|l_i-l_j|},
\end{equation}
there are more non-trivial relations in neutrino sector than in quark sector.
One of them is 
\begin{equation}
(V_{32})^2\sim \frac{m_{\nu 2}^2}{m_{\nu 3}^2}\sim 
\frac{\Delta m_\odot^2}{\Delta m_{atm}^2}\equiv R.
\end{equation}
The ratio $R$ is dependent on the solutions for solar neutrino problem as
\begin{equation}
R\sim \lambda^2({\rm LMA}), \lambda^{3-4}({\rm SMA}), \lambda^6({\rm LMO}),
      \lambda^{11}({\rm VAC}).
\end{equation}
It is obvious that the LMA solution for
solar neutrino problem gives the best value of $V_{32}\sim 0.5$, which
is near the experimental value $(V_{32})\sim 0.7$. This has not been emphasized
in the literature, though we pointed out that LMA solution is naturally 
realized 
in GUT with anomalous $U(1)_A$ symmetry\cite{maekawa,BM} 

Because all the coefficients are determined by the anomalous $U(1)_A$ charges,
the scales of VEVs are also determined by the charges (no flat 
direction) as
\begin{equation}
\VEV{O_i}\sim \left\{ 
\begin{array}{ccl}
  \lambda^{-o_i} & \quad & o_i\leq 0 \\
  0              & \quad & o_i>0
\end{array} \right. , 
\label{VEV}
\end{equation}
where $O_i$ are GUT gauge singlet operators 
~($G$-singlets) with anomalous $U(1)_A$ charges $o_i$.\cite{maekawa,BM,NGCU}
Note that because all positively charged operators have vanishing VEVs,
SUSY zero (holomorphic zero) mechanism act well, namely negatively charged
interactions are forbidden by the anomalous $U(1)_A$ gauge symmetry.
This SUSY zero mechanism plays an important role in solving the doublet-triplet
splitting problem in a natural way, that is not discussed here.

\section{Why M-theory?}
Using anomalous $U(1)_A$ gauge symmetry, we can obtain GUT models in which
doublet-triplet splitting is realized with generic interactions.
However, it is innevitable in the scenario that mass spectrum of
superheavy fields does not respect $SU(5)$ symmetry.
Naively thinking, it spoils the success of gauge coupling unification 
in minimal $SU(5)$ GUT. In the early stage of our works, we expected
that there must be a tuning parameter (charge) because the rank of
$SO(10)$ or $E_6$ is larger than that of $SU(5)$.
However, the fact was more exciting than what we expected.

Because in our scenario, mass spectrum of superheavy fields and VEVs 
(and unification scale $\Lambda_u\sim \lambda^{-a}$) are 
determined by $U(1)_A$ charges,
gauge coupling unification conditions 
$\alpha_1(\Lambda_u)=\alpha_2(\Lambda_u)=\alpha_3(\Lambda_u)$ are
rewritten by the $U(1)_A$ charges as
\begin{eqnarray}
\Lambda&\sim& \Lambda_G, \\
h&\sim& 0.
\end{eqnarray}
It is amazing that all the charges except that of doublet Higgs are
cancelled out in the conditions.\cite{NGCU} It means that there is no
tuning parameters for the gauge coupling 
unification other than those in the minimal $SU(5)$ GUT.
Actually, the first relation defines the scale
of the theory and the second relation corresponds to that for the colored
Higgs mass in the minimal $SU(5)$ GUT because the charge of doublet Higgs 
is the same as the charge of triplet Higgs.

Let us explain the situation. 
For any cutoff and for any charges,
we can calculate the running gauge couplings $\tilde g_i(\Lambda_W)$
at the weak scale from the 
theory which is defined at the cutoff scale, in the GUT with anomalous $U(1)_A$.
The above cancellation means that the three running gauge couplings $g_i(\mu)$,
which are calculated in MSSM by using the gauge couplings 
$\tilde g_i(\Lambda_W)$ as the initial values at the weak scale,
always meet at a scale,
which is nothing but the cutoff scale in the scenario. 
The fact that three gauge 
couplings meet at the usual GUT scale $\Lambda_G\sim 2\times 10^{16}$ GeV
indicates that in the scenario of GUT with anomalous $U(1)_A$ the cutoff
scale must be around the scale $\Lambda_G$. The real GUT scale
is estimated as $\Lambda_U\sim \lambda^{-a}\Lambda_G$, which is smaller than
the usual GUT scale. Therefore, proton decay via dimension 6 operators
is one of the most interesting predictions in our scenario.
If $a=-1$, the lifetime of proton is estimated as
\begin{equation}
\tau_p(p\rightarrow e\pi^0)\sim 1\times 10^{34}
\left(\frac{\Lambda_A}{5\times 10^{15}\ {\rm GeV}}\right)^4
\left(\frac{0.01({\rm GeV})^3}{\alpha}\right)^2  {\rm yrs},
\end{equation}
using hadron matrix element parameter $\alpha$ calculated by 
lattice.\cite{JLQCD}
It is also quite interesting that these results are independent of the details
of the Higgs sector. The sufficient conditions are 
just the following three conditions;
1, Unification group is simple. 2, The VEVs are given by eq. (\ref{VEV}).
3, At a low energy scale, MSSM(+singlets) is realized.

What is important here for this talk is that the cutoff scale must be taken
the usual GUT scale which is smaller than the string scale or Planck scale.
Lowering the string scale is one of the most interesting subjects in the 
string phenomenology, and Horava-Witten pointed out that the extra-dimension
in which only gravity modes can propagate realizes lower string scale in the
context of heterotic M-theory.\cite{HW}

\section{Why heterotic?}
In our scenario, $E_6$(or $E_8$) gauge group plays an important role in
realizing bi-large neutrino mixings and in solving SUSY flavor problem
by non-abelian horizontal symmetry. Therefore, if other type of 
superstring theory can induce $E_6$ (or $E_8$) gauge group, it is also 
considerable.

\subsection{Bi-large neutrino mixings}
First, it is explained that $E_6$ is important in realizing bi-large neutrino
mixings. 
 The fundamental 
representation of $E_6$ is divided as
\begin{equation}
\Psi_i({\bf 27})\rightarrow {\bf 16}_i[{\bf 10}_i+{\bf \bar 5}_i+{\bf 1}_i]
+{\bf 10}_i[{\bf 5}_i+{\bf \bar 5'}_i]+{\bf 1}_i[{\bf 1}_i]
\end{equation}
under $(E_6\supset)SO(10)[\supset SU(5)]$, where $i=1,2,3$ is a generation
index. What is important here is that there are six ${\bf \bar 5}$ fields,
three linear combination of which become massless. This structure gives
different anomalous $U(1)_A$ charges of the massless ${\bf \bar 5}$ fields
from those of ${\bf 10}$ fields, that results in different mixing matrices
of quark and lepton. This is one of the most interesting feature in 
$E_6$ GUT, which Bando and Kugo called ``E-twisting" structure.\cite{BK} 
Unfortunately, they
examined special cases, in which SMA solution for solar neutrino problem
is realized. However, as discussed in the previous section, in many models
using $U(1)$ type Froggatt-Nielsen mechanism, LMA solution is the most natural
solution for the solar neutrino problem.
Moreover,  this ``E-twisting" structure naturally explain that the mixing 
angles of
lepton sector become larger than those of quark sector.
Roughly speaking, this is because ${\bf \bar 5}_3$ 
and ${\bf \bar 5'}_3$ naturally become superheavy because 
of the larger charge of third generation field $\Psi_3$ than
those of the first 2 generation fields.
Let us introduce 
the following interactions
\begin{equation}
\lambda^{\psi_i+\psi_j+\phi}\Psi_i\Psi_j\Phi
+\lambda^{\psi_i+\psi_j+c}\Psi_i\Psi_j C,
\end{equation}
where a non-vanishing VEV $\VEV{\Phi}$ breaks $E_6$ into $SO(10)$ and
gives mass terms of ${\bf 5}_i$ and ${\bf \bar 5}_j$, and
a non-vanishing VEV $\VEV{C}$ breaks $SO(10)$ into $SU(5)$ and 
gives mass terms of ${\bf 5}_i$ and ${\bf \bar 5'}_j$.
It is natural to expect that ${\bf \bar 5}_3$ and ${\bf \bar 5'}_3$
become superheavy because $\psi_3<\psi_2,\psi_1$.
Therefore, main modes of three massless fields ${\bf \bar 5}$ come
from the first two generation fields, for example, 
$({\bf \bar 5}_1,{\bf \bar 5'}_1,{\bf \bar 5}_2)$.
It is obvious that this results in milder hierarchy of ${\bf \bar 5}$ 
charges than that of ${\bf 10}$ charges. Therefore, larger mixing angles
in lepton sector than those in quark sector are naturally obtained because
the mixing angles are obtained as $V_{ij}\sim \lambda^{|c_i-c_j|}$, where
$c_i$ are $q_i$ or $l_i$. And actually, realistic quark and lepton masses
and mixings including bi-large neutrino mixings can be obtained in $E_6$
models as in the model in Table 1. In the model, the main component of
doublet Higgs in MSSM comes from ${\bf 10}_\Phi$ of $SO(10)$ and 
Yukawa couplings of ${\bf \bar 5'}_1+\lambda^\Delta{\bf \bar 5}_3$
are mainly induced via the mixing $\lambda^\Delta{\bf \bar 5}_3$,
where $\Delta\sim 2.5$ is expected in the model in Table 1.
Note that in $E_6$ GUT with anomalous $U(1)_A$, larger neutrino
mixing angles than quark angles are automatically realized, while 
$SO(10)$ GUT scenario\cite{maekawa,NY} can reproduce bi-large neutrino 
mixings but not automatically.

\subsection{Non-abelian horizontal symmetry as SUSY flavor problem}
Second, it is shown that the fact that the massless ${\bf\bar 5}$ fields
come from the first two generation fields plays an important role in 
solving SUSY flavor problem by non-abelian horizontal symmetry.\cite{horiz}

Before explaining the main point, we show that non-abelian horizontal
gauge symmetry can be naturally embedded in the scenario of GUT with 
anomalous $U(1)_A$.\cite{horiz} 
Pick up an example in Table 1. The $SU(2)_H$ breaking
scale is determined by the VEV 
$\VEV{F^a\bar F_a}\sim \lambda^{-(f+\bar f)}$, which is given by the relation
(\ref{VEV}). The VEVs can be taken as 
$\VEV{F^T}=\VEV{\bar F}=(0, \lambda^{-\frac{1}{2}(f+\bar f)})$,
using the $D$-flatness condition of $SU(2)_H$ and $SU(2)_H$ gauge symmetry.
Considering the $SU(2)_H\times U(1)_A$ invariants
\begin{eqnarray}
\lambda^{\phi+f}\Psi^a\VEV{F^b}\epsilon_{ab}&\sim & 
\lambda^{\phi-\frac{1}{2}(\bar f-f)}\Phi_1
\equiv \lambda^{\tilde\phi_1}\Phi_1, \\
\lambda^{\phi+\bar f}\Psi^a\VEV{\bar F_b}&\sim & 
\lambda^{\phi+\frac{1}{2}(\bar f-f)}\Phi_2
\equiv \lambda^{\tilde\phi_2}\Phi_2, \\
\lambda^{\phi_3}\Phi_3&\equiv&\lambda^{\tilde\phi_3}\Phi_3,
\end{eqnarray}
it is obvious that the effective charges $\tilde \phi_i$ $(i=1,2,3)$ 
determine the hierarchy in the superpotential. As a matter of fact, 
we can almost forget $SU(2)_H$ in considering the hierarchy of Yukawa 
couplings in the superpotential. Therefore, the discussion in the previous 
subsection
is applicable even with non-abelian horizontal gauge symmetry.
On the other hand, in the K\"ahler potential, $SU(2)_H$ symmetry leads
to degenerate scalar fermion masses of the first two generation fields.
The scaler fermion mass matrices of the model in Table 1 can be estimated
as
\begin{equation}
\frac{\tilde m^2_{\bf 10}}{\tilde m^2}\sim
\left(\matrix{1+\lambda^4 & \lambda^5 & \lambda^3 \cr
              \lambda^5 & 1+\lambda^4 & \lambda^2 \cr
              \lambda^3 & \lambda^2 & O(1) }\right),\quad
\frac{\tilde m^2_{\bf \bar 5}}{\tilde m^2}
\left(\matrix{1+\lambda^4 & \lambda^6 & \lambda^5 \cr
              \lambda^6 & 1+\lambda^2 & \lambda^7 \cr
              \lambda^5 & \lambda^7 & 1+\lambda^4 }\right)
\end{equation}
where the corrections come from the interactions including $F$ or $\bar F$, 
and the spurion field $Z$, which 
has non-vanishing $F$-term component 
$\VEV{F_Z}=\tilde m\Lambda$, such as $K=|\Psi^a\bar F_a|^2Z^\dagger Z$.
It is quite impressive that all the three generation ${\bf\bar 5}$ fields 
have degenerate scalar fermion masses in leading order because they come 
from the first two generation fields. This feature is important in 
suppressing FCNC. 
The various FCNC processes constrain the mixing matrices defined by
$\delta_{\bf R}\equiv V_{\bf R}^\dagger \Delta_{\bf R}V_{\bf R}$
\cite{GGMS} (${\bf R}={\bf 10}$ or ${\bf \bar 5}$), where  
$\Delta_{\bf R}\equiv (\tilde m^2_{\bf R}/\tilde m^2)-1$, 
$V_{\bf 10}\sim V_{CKM}$ 
and $V_{\bf \bar 5}\sim V_{MNS}$. Here $V_{CKM}$ is the 
 Cabbibo-Kobayashi-Maskawa matrix and $V_{MNS}$ is the Maki-Nakagawa-Sakata
 matrix.
In this model, these
mixing matrices are approximated as  
\begin{equation}
\delta_{\bf 10}=\left(\matrix{\lambda^4 & \lambda^5 & \lambda^3 \cr
                          \lambda^5 & \lambda^4 & \lambda^2 \cr
                          \lambda^3 & \lambda^2 & 1 }
             \right), \ 
\delta_{\bf \bar 5}=
          \left(\matrix{\lambda^3 & \lambda^{2.5} & \lambda^3 \cr
                          \lambda^{2.5} & \lambda^2 & \lambda^{2.5} \cr
                          \lambda^3 & \lambda^{2.5} & \lambda^3 }
             \right)
\label{delta}
\end{equation}
at the GUT scale.  The constraints at the weak scale from 
$\epsilon_K$ in $K$ meson mixing,
\begin{eqnarray}
\sqrt{|{\rm Im}(\delta_{d_L})_{12}(\delta_{d_R})_{12})|}&\leq&
 2\times 10^{-4}\left(\frac{\tilde m_q}{500 {\rm GeV}}\right) \\
|{\rm Im}(\delta_{d_R})_{12}| & \leq & 1.5\times 10^{-3}
\left(\frac{\tilde m_q}{500 {\rm GeV}}\right),
\end{eqnarray}
requires scalar quark masses larger than
500 GeV, because in this model
$\sqrt{|(\delta_{d_L})_{12}(\delta_{d_R})_{12})|}\sim 
\lambda^{7.5}(\eta_q)^{-1}$
and $|(\delta_{d_R})_{12}|\sim \lambda^{2.5}(\eta_q)^{-1}$, where
we take a renormalization factor $\eta_q\sim 6$.\footnote{
The renormalization factor is strongly dependent on the ratio of the gaugino 
mass to the scalar fermion mass and the model below the GUT scale.
If the model is MSSM and the ratio at the GUT scale is 1, 
then $\eta_q=6\sim 7$.}
And the constraint from the $\mu\rightarrow e\gamma$ process, 
\begin{equation}
|(\delta_{l_L})_{12}|\leq 4\times 10^{-3}\left(
\frac{\tilde m_l}{100 {\rm GeV}}\right)^2,
\end{equation}
requires scalar lepton masses larger than 200 GeV,
because $|(\delta_{l_L})_{12}|\sim \lambda^{2.5}$ in this model.

Note that in $SO(10)$ GUT with three ${\bf 16}$ and one 
${\bf 10}$,\cite{maekawa}
in which one of the massless ${\bf \bar 5}$ fields comes from
the ${\bf 10}$ of $SO(10)$, non-abelian horizontal symmetry does not 
guarantee the degenerate scalar fermion masses of
three massless ${\bf \bar 5}$ fields.

\section{Discussion and summary}
$E_6$ unified group is important in explaining larger mixings in lepton
sector than in quark sector. Moreover, in $E_6$ GUT, introducing 
non-abelian horizontal gauge symmetry $SU(2)_H$ or $SU(3)_H$ can solve 
SUSY flavor problem even with bi-large neutrino mixings. 
In the resulting models, all the three generation quarks and leptons are
unified into one or two multiplets, which is important in solving SUSY
flavor problem.\footnote{
Strictly speaking, the subgroup $SU(2)_E$ of $E_6$, which rotates
two ${\bf \bar 5}$ fields in ${\bf 27}$ of $E_6$, is sufficient
to realize this interesting structure. Therefore, the unified gauge
group can be $SU(6)\times SU(2)_E$, $SU(3)^3$ or $SO(10)'\times U(1)$
\cite{flipped}. The trinication $SU(3)^3$ is interesting because
doublet-triplet splitting is realized,\cite{shafi} 
but this is not a simple group,
so natural gauge coupling unification is not guaranteed.}
 Therefore, $E_8$ group looks promizing because
$E_8$ has a maximal subgroup $E_6\times SU(3)_H$,
though $E_8$ GUT cannot be realized in 4 dimensional theory because
$E_8$ has no chiral representation.

In the scenario of GUT with anomalous $U(1)_A$ symmetry, the fact that
in MSSM three gauge couplings meet at a scale $\Lambda_G$ means that
the cutoff scale of the theory must be the scale $\Lambda_G$. 
Because the scale $\Lambda_G$ is smaller than the Planck scale,
this may suggest that there are extra dimensions in which only gravity
modes can propagate, as discussed by Horava-Witten.

Considering above two facts, GUT with anomalous $U(1)_A$ must imply
hoterotic M-theory.

Of course it is known that in weakly coupled heterotic string with
Kaz-Moody level 1 cannot produce adjoint Higgs fields, which are
required in our scenario. However, considering level 2 models,
there are some models with $E_6\times SU(2)_H$ with two $E_6$
adjoints.\cite{erler} Though these models do not include
anomalous $U(1)_A$ gauge symmetry, the existence is quite impressive.

We hope that our study becomes a breakthrough in string phenomenology,
and in near future, a path from the superstring to the
standard model is discovered.

\section*{Acknowledgments}
The author thanks T. Yamashita, M. Bando and Q. Shafi for their 
collaborations and stimulating discussions. And also thanks T. Kugo, 
T. Kobayashi and J. Erler for interesting discussions. 
N.M. is supported in part by 
Grants-in-Aid 
for Scientific Research from the Ministry of Education, Culture, Sports, 
Science and Technology of Japan.

\end{document}